\documentclass[12pt,draftclsnofoot,onecolumn]{IEEEtran} 
\usepackage{graphicx,cite,enumitem,}             		
			
\usepackage{amssymb,amsfonts,amsthm}   

\usepackage[cmex10]{amsmath}
\usepackage[unicode,colorlinks, linkcolor=black,anchorcolor=black,citecolor=black]{hyperref}

\newtheorem{theorem}{Theorem}

\newtheorem{definition}{Definition}

\begin{document}

\title{Semidefinite Relaxation Based Blind Equalization using Constant Modulus Criterion}

\author{Kun~Wang,~\IEEEmembership{Member,~IEEE,}
        and~Zhi~Ding,~\IEEEmembership{Fellow,~IEEE}
\thanks{K. Wang is with Qualcomm, Inc, Santa Clara, CA 95051. Email: kunwang@ieee.org.}
\thanks{Z. Ding is with the Department of Electrical and Computer Engineering, University of California, Davis, CA 95616, USA. Email: zding@ucdavis.edu}
}

\maketitle

\begin{abstract}
Blind equalization is a classic yet open problem.
Statistic-based algorithms, such as constant modulus (CM), were widely investigated.
One inherent issue with blind algorithms is the phase ambiguity of equalized signals.
In this letter, we propose a novel scheme based on CM criterion and 
take advantage of the asymmetric property in a class of LDPC codes to resolve the phase ambiguity.
Specifically, a new formulation with modified CM objective function and relaxed code constraints 
is presented. 
\end{abstract}


\section{Introduction}
In digital and especially wireless communications, the channels introduce distortions that can hamper accurate signal recovery at receivers. In particular, the single-input single-output (SISO) dispersive channels can incur inter-symbol interference (ISI). Besides ISI, the signals transmitted in the multi-input multi-output (MIMO) systems suffer from co-channel interference (CCI) from other data streams. Both of them require channel equalization at the receiver to avoid errors in data detection \cite{qureshi1985adaptive,proakisdigital}. Among various channel equalization methods, linear equalization admits the simplest form by applying feed-forward linear filters to compensate the channel distortions. Typically, the transmitters need to insert known pilot symbols within data frames for channel estimation or equalizer training. Nonetheless, pilot symbols are not available in some scenarios and also training decreases the overall system throughput. By eliminating training data and maximizing channel capacity for data-bearing transmissions, blind channel equalization presents a bandwidth efficient solution to the distortion compensation.

There has been a plethora of research papers on blind channel equalization and related topics, e.g., blind source separation (BSS); cf. \cite{ding2001blind}. A vast amount of early works are done under the framework of statistical analysis. Specifically, a lot of blind equalization schemes are achieved by exploiting second-order \cite{tong1994blind,abed1997prediction} or higher-order statistics (HOS) \cite{mendel1991tutorial,chi2003batch}. These statistics based algorithms, such as the well-known constant modulus algorithm (CMA) \cite{johnson1998blind} and the super-exponential algorithm (SEA) \cite{shalvi1993super}, directly minimize special non-convex cost functions, and thus tend to exhibit local convergence\cite{ding1992whereabouts,li1995convergence}. Even though HOS algorithms may provide satisfactory performance in a certain cases, a relatively large number of samples are needed due to the nature of higher-order statistics. This drawback limits their applications when the environment is fast time-varying.

As recent advancement in the field of optimization, many non-convex and NP-hard quadratically constrained quadratic program (QCQP) can be reformulated as semidefinite program (SDP) with the rank-1 relaxation, which leads to the so-called semidefinite relaxation (SDR)\cite{luo2010semidefinite,palomar2010convex}. 
To tackle the issues of local convergence and large sample requirements, this SDR technique is widely used in detection and equalization, since it can generate a provably approximately optimal solution with a randomized polynomial time complexity. 
The SDR MIMO detection in BPSK and QPSK shows promising performance \cite{wang2018non,tan2001application,wang2018iterative} and are extended to other constellations \cite{ma2004semidefinite,mobasher2007near}. Specifically, blind channel equalization, blind source separation and blind MIMO detection are treated in \cite{li2001blind,li2003blind,ma2006blind}, respectively. In \cite{li2001blind}, the authors take advantage of the minimum mean square error (MMSE) criterion to formulate the blind channel equalization problem into a quadratic optimization with binary constraints. Later on, this idea is extended to BSS problem in \cite{li2003blind} with the exploitation of known input alphabets. On another front, efficient high-performance implementations for blind maximum-likelihood (ML) detection of orthogonal space-time block code (OSTBC) is investigated in \cite{ma2006blind}.

In this work, we intend to utilize the \emph{constant modulus} (CM) criterion. Scanning the literature, we note that a similar work by Mariere \textit{et.~al} converts the original CM cost function into a SDP by equating the corresponding polynomial coefficients \cite{mariere2003blind}. The resulting SDP is of high notational and computational complexities, and requires an alternating projection algorithm as post-processing to compute the final equalizer vector. To deal with the complexity, we modify the original CM objective function by changing the $\ell_2$ norm into $\ell_1$ norm, which leads to a substantially simpler form of SDP. Moreover,
in the blind equalization community, it is generally acknowledged that scalar multiplicative and phase rational ambiguities are inherent in the equalized symbols. What's more, the blind algorithms may recover a different data stream in MIMO detection, although the prior information of the interested stream is provided \cite{chi2003batch}. 
The aforementioned algorithms including both HOS-based and optimization-based methods, however, do not address these critical issues.
The scalar multiplicative issue is relatively easy to solve by rescaling the power of the equalizer output, whereas the phase ambiguity is quite challenging since almost all the blind cost functions are insensitive to the phase rotations. The usual way for fixing phase is to utilize a reference symbol \cite{zia2010linear}, which essentially also reduces the information data rate. 
Since most current wireless systems already employ forward error correction (FEC) codes, we plan to exploit the information embedded in the FEC code (LDPC code in our case). One noticeable property of \emph{asymmetric} LDPC code is that the negation of any valid codeword does not belong to the code \cite{scherb2003phase,scherb2005phase}.
To take advantage of the code information, we will use the relaxed code constraints in real/complex domain \cite{feldman2005using}.
Actually, this set of code constraints have been widely used in our works: 
space-time code is concatenated with LDPC code in \cite{wang2014joint,wang2015joint}; 
partial channel information is treated in \cite{wang2015diversity,wang2016diversity};
multi-user scenario with different channel codes or different interleaving patterns are handled in \cite{wang2016robust,wang2016diversity};
SDR-based MIMO detector is proposed in \cite{wang2018integrated} that can approach ML performance.
Different from previous works \cite{wang2017galois}, the asymmetry property will be explored in this work to resolve the phase ambiguity.

\section{SDP Formulations}

Here we consider a transmission that takes the form
\begin{equation}
\mathbf{x}[n] = \mathbf{H} \mathbf{s}[n] + \mathbf{v}[n], \quad 1 \leq n \leq N
\end{equation}
where $\mathbf{H}$ is the (equivalent) communication channel, $\mathbf{s}[n]$ is the transmitted signal, $\mathbf{x}[n]$ is the received signal, and $\mathbf{v}[n]$ is the additive white Gaussian noise (AWGN) at the receiver. These vectors and matrix are of appropriate size. Note that this system model is quite general in the sense that it incorporates spatial multiplexing MIMO, space-time coded MIMO in the \emph{equivalent spatial diversity} model, and SISO transmission in frequency-selective channel with $\mathbf{H}$ being a Toeplitz matrix.

To begin with the algorithm development, we denote the desired equalizer vector by $\mathbf{w}$. In MIMO detection, this equalizer is aimed for a certain stream and length of the equalizer is equal to the number of receive antennas; for SISO ISI equalization, the equalizer length is related to the channel delay spread. Let $y[n]$ be the equalized symbol, that is, $y[n] = \mathbf{w}^H \mathbf{x}[n]$ for flat-fading MIMO channel and $y[n] = w[n] \ast x[n]$ for SISO ISI channel.
In the sequel, we only consider the SISO ISI channel for derivation simplicity. Without loss of generality, assume the equalizer is $(L+1)$ in length. In order to have a compact form for SISO ISI case, denote $\mathbf{x}_n = [x_n, x_{n-1}, \ldots, x_{n-L}]^T$ and thus $y[n] = \mathbf{w}^H \mathbf{x}_n$.

\subsection{Basic CM-based SDP Formulation}
If perfect equalization is achieved, the sequence $y[n]$ will be of the same modulus as channel input signal $s[n]$. For BPSK or QPSK, the modulus of $y[n]$ is expected to be 1 and consequently a natural formulation of the blind CM equalization is 
\begin{equation} \label{CM_cost}
\text{min.} \quad \mathbf{J}(\mathbf{w}) 
= \frac{1}{N} \sum_{n=1}^{N} \left| |y[n]|^2 - 1\right| 
= \frac{1}{N} \sum_{n=1}^{N} \left| \mathbf{w}^H \mathbf{X}_n \mathbf{w} - 1 \right| 
\end{equation} 
where the second equality follows
\begin{equation}
\vert y[n] \vert ^2 = \mathbf{w}^H \mathbf{x}_n \mathbf{x}_n^H \mathbf{w} = \mathbf{w}^H \mathbf{X}_n \mathbf{w}
\end{equation}

By introducing auxiliary variable $\tau_n$, we can transform the unconstrained problem (\ref{CM_cost}) to the following constrained problem
\begin{equation} \label{cm_qcqp}
\begin{aligned}
& \underset{\mathbf{w}}{\text{min.}}
& & \frac{1}{N} \sum_{n=1}^N \tau_n \\
& \text{s.t.}
& & \mathbf{w}^H \mathbf{X}_n \mathbf{w} - \tau_n \leq 1, \; 1 \leq n \leq N \\
& 
& & \mathbf{w}^H \mathbf{X}_n \mathbf{w}  + \tau_n \geq 1, \; 1 \leq n \leq N
\end{aligned}
\end{equation}
where $\mathbf{w}$ and $\tau_n$'s are optimization variables. However, notice that the second quadratic constraints do not define a convex set, and therefore this QCQP is non-convex. Let $\mathbf{W} = \mathbf{w} \mathbf{w}^H$ and then this QCQP can be transformed into a convex SDP without the rank-1 requirement of $\mathbf{W}$, shown as follows
\begin{equation} \label{cm_sdp}
\begin{aligned}
& \underset{\mathbf{w}}{\text{min.}}
& & \frac{1}{N} \sum_{n=1}^N \tau_n \\
& \text{s.t.}
& & \text{tr} (\mathbf{X}_n \mathbf{W}) - \tau_n \leq 1, \; 1 \leq n \leq N \\
& 
& & \text{tr} (\mathbf{X}_n \mathbf{W})  + \tau_n \geq 1, \; 1 \leq n \leq N
\end{aligned}
\end{equation}
where $\mathbf{W}$ and $\tau_n$'s are optimization variables. Upon obtaining the optimal solution $\mathbf{W}^*$, rank-1 approximation or randomization can be used to find the desired equalizer $\mathbf{w}^*$ in $\epsilon$-accuracy \cite{ye1999approximating}.

\subsection{Integration of LDPC Code Constraints}
The linear programming decoding proposed by Feldman \textit{et.~al} opens a gateway for the unification of the detection and decoding processes \cite{feldman2005using}. 
Specifically, Consider an LDPC code $\mathcal{C}$. Let $\mathcal{M}$ and $\mathcal{N}$ be the set of check nodes and variable nodes of the parity check matrix $\mathbf{H}$, respectively. 
Denote the set of neighbors of the $m$-th check node as $\mathcal{N}_m$. For a subset $\mathcal{F} \subseteq \mathcal{N}_m$ with odd cardinality $|\mathcal{F}|$, the explicit constraints on the coded bits $f[n]$ are given by the following parity check inequalities 
\begin{equation} \label{eq:parity_ineq}
\sum_{ n \in \mathcal{F} } f[n] - \sum_{ n \in (\mathcal{N}_m \backslash \mathcal{F})} f[n] \leq |\mathcal{F}| - 1, \quad \forall m \in \mathcal{M}, \mathcal{F} \subseteq \mathcal{N}_m, |\mathcal{F}| \, \text{odd}
\end{equation}
and box constraints
\begin{equation} \label{eq:box_ineq}
0  \leq f[n] \leq 1, \quad \forall n \in \mathcal{N}.
\end{equation}

It is worthwhile to note that the negation of a valid codeword may still be valid for a generic LDPC code $\mathcal{C}$, and thus the above code constraints cannot fix the phase rotations. However, one special class of LDPC code with the asymmetry property can help to prevent such bad configurations. Its definition is stated in \cite{scherb2003phase} and repeated here. 
\begin{definition}
A channel code is called asymmetric if the negation of an arbitrary valid codeword is not a valid codeword, i.e. $\mathbf{c} \in \mathcal{C} \Rightarrow \overline{\mathbf{c}} \notin \mathcal{C}$.
\end{definition}

Obviously, a code is asymmetric if an arbitrary parity check sum node includes an odd number of neighbors, that is, $| \mathcal{N}_m |$ is odd for every $m \in \mathcal{M}$. This observation is formalized in the following theorem \cite{scherb2005phase}.
\begin{theorem}
If there exists an arbitrary row or an arbitrary linear combination of rows in $\mathbf{H}$ such that the number of 1's is odd, then this code is asymmetric.
\end{theorem}

However, to further integrate the LDPC code constraints, we need to explicitly have the variable $\mathbf{w}$ in the optimization problem. Ideally, we want to impose the constraint $\mathbf{W} = \mathbf{w} \mathbf{w}^H$. Nonetheless, this constraint is not convex. As inspired by \cite{vandenberghe1996semidefinite}, we approximate the exact constraint by a convex constraint in the form $\mathbf{W} \succeq \mathbf{w} \mathbf{w}^H$, which is equivalent to
\begin{equation} \label{eq:psd}
\begin{bmatrix}
\mathbf{W} & \mathbf{w} \\
\mathbf{w}^H & 1
\end{bmatrix}
\succeq 0.
\end{equation}

The last step for this integration is to use the squeezing box constraints and symbol-to-bit mapping constraints. In particular, for BPSK, the two kinds of constraints are
\begin{equation} \label{eq:squeeze}
\left| \mathbf{w}^H \mathbf{x}_n - z[n] \right| \leq t_n, \quad 1 \leq n \leq N
\end{equation}
and
\begin{equation} \label{eq:mapping}
z[n] = 2 f[n] - 1, \quad 1 \leq n \leq N
\end{equation}
where $z[n]$ is a dummy variable which represents the point on data constellation, and $t_n$ is the variable to be lifted into the cost function for squeezing box. For simplification, the constraints (\ref{eq:psd}), (\ref{eq:squeeze}) and (\ref{eq:mapping}) can be categorized as connection constraints.

To this end, the SDP with code constraints for fixing phase rotation is as follows
\begin{equation} \label{cm_code_sdp}
\begin{aligned}
& \underset{\mathbf{w}}{\text{min.}}
& & \frac{1}{N} \sum_{n=1}^N \tau_n + \sum_{n=1}^N t_n \\
& \text{s.t.}
& & \text{tr} (\mathbf{X}_n \mathbf{W}) - \tau_n \leq 1, \; 1 \leq n \leq N \\
& 
& & \text{tr} (\mathbf{X}_n \mathbf{W})  + \tau_n \geq 1, \; 1 \leq n \leq N \\
& 
& & [\text{Connection Constraints (\ref{eq:psd}), (\ref{eq:squeeze}) and (\ref{eq:mapping})}] \\
& 
& & [\text{LDPC Code Constraints (\ref{eq:parity_ineq}) and (\ref{eq:box_ineq})}]
\end{aligned}
\end{equation}

\section{Summary}
In this letter, we first reformulate the non-convex CM cost function into a convex SDP with rank-1 relaxation.   
We further attempt to address the inherent issue of phase rotation by integrating LDPC code constraints.
It is a novel way and a seemingly promising approach of using the prior information embedded in asymmetric LDPC code instead of reference symbol. 
The remaining works are to extend this formulation to higher-order modulations, which is non-trivial, and to exploit other a priori information in the transmission. Preliminary tests are yet to come.

\ifCLASSOPTIONcaptionsoff
  \newpage
\fi



%

\bibliographystyle{IEEEtran}
\bibliography{IEEEabrv,mybibfile}

\end{document}